\documentstyle[epsfig,sprocl]{article}

\def\Journal#1#2#3#4{{#1} {\bf #2}, #3 (#4)}

\def\NPB{{\em Nucl. Phys.} B}
\def\PLB{{\em Phys. Lett.}  B}
\def\PRL{\em Phys. Rev. Lett.}
\def\PRD{{\em Phys. Rev.} D}
\def\ZPC{{\em Z. Phys.} C}

\def\be{\begin{equation}}
\def\ee{\end{equation}}
\def\bea{\begin{eqnarray}}
\def\eea{\end{eqnarray}}

\title{QCD and Total Cross-sections~\footnote{Presented by G. Pancheri at
the XXIX International Symposium on Multiparticle Dynamics, June 1999, Brown
 University, U.S.A. To appear in the Proceedings.}}
\author{Rohini M. Godbole}

\address{Centre for Theoretical Studies, Indian Institute of Science,
Bangalore, 560012,India.}

\author{A. Grau}

\address{Departamento de F\'\i sica Te\'orica y del Cosmos,
Universidad de Granada, 18071 Granada, Spain}

\author{G.Pancheri}

\address{Laboratori Nazionali di Frascati dell'INFN, Via E. Fermi 40,
I00044 Frascati, Italy}
\begin{document}
\begin{flushright}
IISc-CTS/8/99\\
UG-FT-105/99\\ 
LNF-99/030(P)\\
hep-ph/9912395\\
\end{flushright}

\maketitle
\abstracts{ We discuss models for total cross-sections, show 
their predictions for photon-photon collisions and compare them with 
the recent LEP measurements. We show that the extrapolations
to high center of mass energies within various models differ by large factors
at high energies and discuss the precision required from future measurements
 at the 
proposed Linear Collider which would allow to distinguish between them.}

\begin{figure}[htb]
\begin{center}
\mbox{\epsfig{file=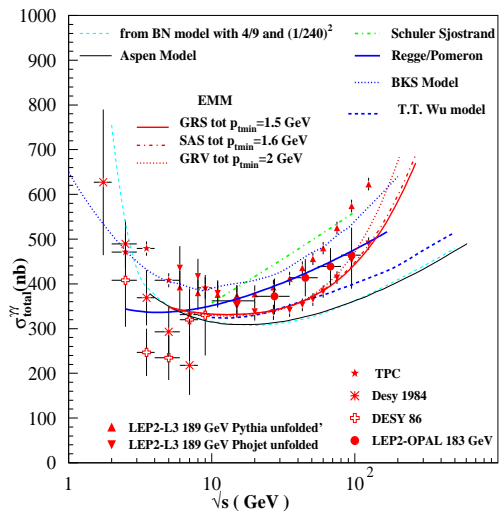, height=60mm,width=80mm}}
\vspace{0.5cm}
\caption{\label{fig:alltogether}Models and data for total 
$\gamma \gamma$ cross-sections}
\end{center}
\end{figure}

In this talk, we shall discuss total cross-sections and the 
contribution to them from QCD processes, both for hadronic and 
photonic reactions. It is by now accepted that
it is possible to `predict' the rising trend of total cross-sections, 
albeit still not with very high accuracy, in a QCD based framework using
phenomenological inputs, particularly through the use of the minijet 
model~\cite{halzen}. This model ascribes the rise of the total cross-sections 
to the increasing number of low $p_T$ partons, and their collisions. 
The present 
knowledge of parton densities in the hadrons has now been extended to
photons by the studies of the resolved photon processes, 
at HERA and LEP~\cite{roh,butterworth}.
Recently measurements of total cross-sections for photonic processes have 
become available from HERA~\cite{ZEUS,H1} and LEP~\cite{L31,L32,OPAL1,OPAL2}.
This is a very important input to the phenomenological efforts  towards 
developing  a realistic model for calculation of total cross-sections. 
Unfortunately, the uncertainties plaguing the experimental measurements 
do not yet allow us to distinguish  between different theoretical models
that are available.  The situation is illustrated in Fig.~\ref{fig:alltogether}
where data from L3 Collaboration~\cite{L31,L32} correspond to two different 
Montecarlo extrapolations, PYTHIA and Phojet, while OPAL~\cite{OPAL1,OPAL2} 
data have been obtained by averaging between them.
The differences between the
predictions of the various theoretical models are even larger.
In  the following we shall describe briefly the models which differ the 
most and indicate with  what precision 
$\sigma^{\rm had}_{\gamma \gamma}$ need be measured in future experiments, 
in order to distinguish between the models.

\begin{figure}
\begin{center}
\mbox{\epsfig{file=
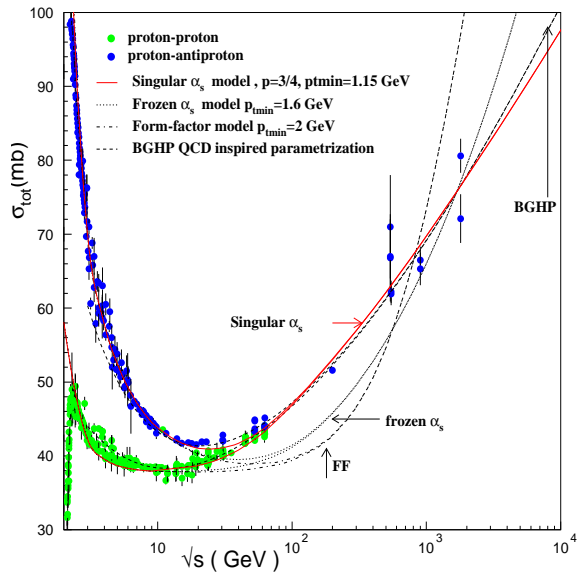,height=60mm,width=80mm}}
\vspace{1cm}
\caption{\label{fig:pptot} Total $p-p$ and ${\bar p}-p$ cross-sections 
compared 
with models from Ref.[12]}
\end{center}
\end{figure}

\begin{figure}[t]
\begin{center}
\mbox{\epsfig{file=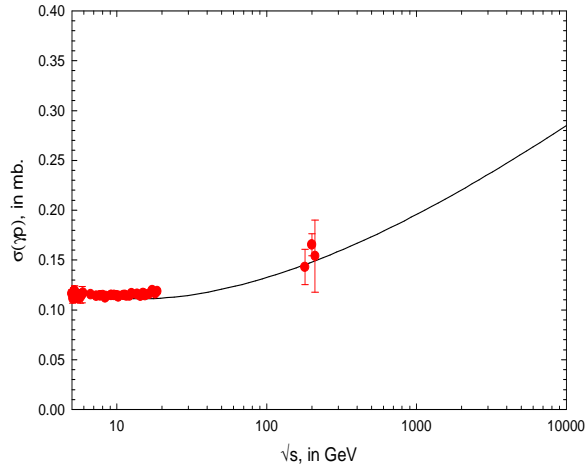,width=80mm,height=60mm}}
\caption{\label{fig:aspengp}Photoproduction total cross-section and the 
Aspen Model$^{10}$ predictions}
\end{center}
\end{figure}

We shall start with the Aspen model~\cite{aspen},
which predicts the photon-photon cross-section starting from 
the QCD inspired model for proton-proton and proton-antiproton 
total elastic
and inelastic cross-sections~\cite{block1}. In this model,
 total cross-sections are obtained through the eikonal 
representation in impact parameter space, i.e.
\begin{equation}
\sigma_{tot}(s)=2\int d^2{\vec b}[1-e^{i\chi(b,s)}]
\label{eq:eikonal}
\end{equation}
with 
the eikonal function parametrized 
through a sum of QCD inspired terms of the type
\begin{equation}
\chi_{ij}(b,s)=W(b, \mu_{ij}) \sigma_{ij}(s)
\end{equation}
with $W(b,\mu_{ij})$ describing the impact parameter space
distribution of partons in the proton obtained as 
\begin{equation}
W(b,\mu_{ij})=\int {{d^2 {\vec q}}\over{(2\pi)^2}}[{\cal F}(q,\mu_{ij})]^2
\end{equation}
where ${\cal F}(q,\mu_{ij})$ is the proton form factor with scale $\mu_{ij}$.
Details about the parametrization of the elementary 
cross-sections $\sigma_{ij}$ can be found in Ref.~[10],
 here we
mention that the functional form reflects the  low x-behaviour of
gluon and quark densities in the protons. The corresponding fit for
proton and antiproton-proton cross-sections  is indicated as the BGHP curve
in Fig.~\ref{fig:pptot}, which we reproduce~\cite{bn2}.
It is based on 11  parameters which allow for a complete description of
the proton-proton and proton-antiproton system, including elastic, total and
differential cross-section, $\rho$-parameter and  nuclear slope. 
One can now describe the photoproduction processes with just two new
inputs, namely Vector Meson Dominance (VMD) and Additive Quark Model
(AQM). This is achieved by using, for the extension to photonic processes,
the expression
\begin{equation}
\label{eq:eikphoton}
\sigma_{tot}^{\gamma p}=2 P_{had} \int d^2{\vec b}
[1-e^{-\chi_I(b,s)}cos{\chi_R}]
\end{equation}
where the 
Vector Meson Dominance factor 
\begin{equation}
\label{eq:phad}
P_{had}=\sum_{V=\rho,\omega,\phi}{{4\pi\alpha}\over{f^2_V}}
\approx {{1}\over{240}}
\end{equation}
with
\begin{equation}
f_\rho=5.64, \ \ {{f_\rho}\over{f_\omega}}={{1}\over{3}}, \ \ 
{{f_\rho}\over{f_\phi}}={{-\sqrt{2}}\over{3}} 
\end{equation}
and 
$\alpha$ evaluated at the $M_Z$ scale. The eikonal is obtained from the
even part
of the corresponding function for proton case, through two simple
AQM inspired substitutions, that is by 
scaling of the s-dependence in the elementary cross-sections as
$\sigma_{ij}^{\gamma p}=2/3 \sigma_{ij}^{p p}$, 
and in the b-shape, i.e. $(\mu_{ij}^{\gamma p})^2=3/2
 (\mu_{ij}^{p p })^2$.  The comparison of the corresponding prediction  with 
the HERA data~\cite{ZEUS,H1} is shown in Fig.~\ref{fig:aspengp}. 
Basically, photoproduction data suggest the value of the parameter $P_{had}$, 
which can then be used, through factorization,
to make a prediction for $\gamma \gamma$ cross-sections. The 
curve predicted by the Aspen model is the lowest one shown in 
Fig.~\ref{fig:alltogether},
and it almost coincides with one obtained by simply scaling the prediction
for proton-proton in the Soft Gluon Summation model of Ref. [12].
The two curves will ultimately differ. They coincide here basically because
the eikonal function is still small and the integrand  
can be expanded. That simple factorization of the
proton-proton  and proton-photon cross-section could give the correct order of
magnitude for  photon-photon case, has been
known~\cite{terazawa} since the first such measurements were
discussed at electron-positron colliders. Interest in this   was recently
rekindled, as is shown by one of the curves indicated in 
Fig.~\ref{fig:alltogether}, {\it i.e.} the one  by T.T. Wu and
collaborators~\cite{ttwu}. The fact that these curves, although
starting from very similar hypothesis, differ when the final
 scaling is applied to photon-photon case, can be ascribed to the fact 
that in this case all the quantities appear squared and 
small differences in the fits to proton-proton and proton-photon processes 
get amplified. All these models, to some extent, consider the
photon to be similar to the proton, and use factorization. 
There is also another popular model, the Regge-Pomeron model, which  
though different in mathematical formulation, belongs to the same
general philosophy.  This model describes the initial decrease and 
the subequent rise as due to the exchange of
different sets of graphs, known as Regge and Pomeron graphs.
Then the cross-section, whose formulation is
based on using analyticity and unitarity is written as
\begin{equation}
\sigma_{tot}=Xs^{\epsilon}+Ys^{-\eta}
\end{equation}
Using a universal set of parameters
for 
Regge and Pomeron trajectories, and factorization at the residues, from
 proton-proton and photo-production~\cite{DL}, one can make the prediction for 
photon-photon shown in Fig.~\ref{fig:alltogether}.
This curve lies higher than most, and in particular than
the one of the Aspen model, but
rises less than the ones from the minijet model, which will be described next.

\begin{figure}[htb]
\begin{center}
\mbox{\epsfig{file=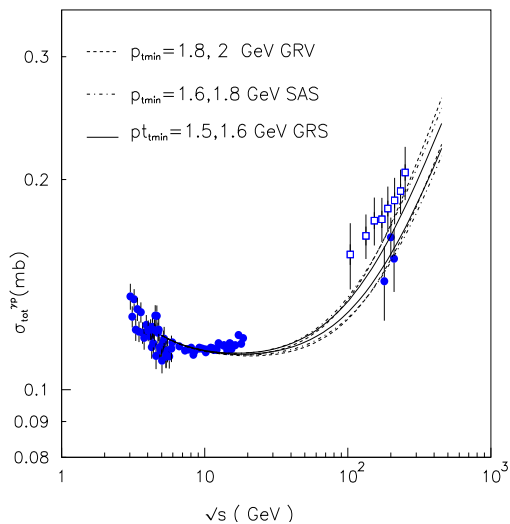,width=80mm,height=60mm}}
\vspace{1cm}
\caption{Photoproduction and extrapolated DIS data in comparison with curves from
the EMM model for different parton densities and $p_{tmin}$.
\label{fig:emmgp}}
\end{center}
\end{figure}

The mini-jet model uses actual parton densities in protons and photons
to describe the rise of the cross-sections. This model 
is unitarized~\cite{durand,halzen2} through the eikonal formulation of
eqs.(\ref{eq:eikonal},\ref{eq:eikphoton}) and one can make predictions for 
photon-photon
starting from photo-production. In order to test the role played
by the QCD jet cross-section, the EMM uses a very simplified form of
the  eikonal
function, which is approximated by a purely imaginary term,  {\it i.e.} 
$\chi_R=0$ and $\chi_I=n(b,s)/2$, with $n(b,s)$ given by the average number 
of collisions at an impact parameter $b$. In the Eikonal Minijet Model (EMM), 
the average number $n(b,s)$ is schematically divided into a soft and a hard
component, i.e.
\begin{equation}
\label{eq:nbs}
n(b,s)=n_{soft}(b,s)+n_{hard}(b,s)
\end{equation}
with the soft term parametrized so as to reproduce
the low energy behaviour of the cross-section, and
\begin{equation}
\label{eq:nhard}
n_{hard}(b,s)=A(b)\sigma_{jet}(s,p_{tmin}) /P_{had}
\end{equation}
where $A(b)$ is obtained by  convoluting the
electromagnetic form factors of the colliding
particles. For the photon, one simple possibility is to use the pion
form factor, identifying the photon as just a $q{\bar q}$-pair, 
for this purpose.  Another possibility is to use Fourier transform of the
intrinsic transverse momentum distribution~\cite{emm}(IPT).
The jet cross-section depends upon the specific set of parton densities,
and, because of the Rutherford singularity, crucially changes according 
to the lowest cut-off used in the calculation, namely $p_{tmin}$. 
Presently different sets of photon densities
are available and predictions can differ. We show in Fig.~\ref{fig:emmgp} the
dependence of the predictions of the Eikonal Minijet Model (EMM) on the
photonic parton densities for three different available sets {\it viz.}, 
GRV~\cite{GRV} , SAS~\cite{SAS} and  GRS~\cite{GRS} as well as on the
$p_{tmin}$. 
The extension to the photon-photon system then proceeds as described in
Ref. [18]
and the outcome is shown in Fig.~\ref{fig:alltogether} as the three
curves  which rise faster than all the others. 

To partly understand the difference in
predictions between these
different models, one can look
at the EMM and Aspen model, which use the same eikonal formulation, 
are both inspired by QCD in the energy dependence and have
similar, albeit not identical, impact parameter description of the collision.
The difference between the Aspen model and the EMM is mostly to be
ascribed to the use of actual parton densities in the jet cross-section
in the latter.
Indeed, the Aspen model starts with a successful parametrization of the
proton case and moves through factorization to describe photon processes,
whereas the EMM describes photon-photon collisions basically by using 
only  the photoproduction and extending the photon properties deduced from
photoproduction, namely scaling the soft part  using AQM and VMD to the
$\gamma \gamma$ case.
Were one to make a straightforward application of the mini-jet model to the 
proton-proton case, as shown in~Ref. [12] and indicated by the curve 
labelled FF in Fig.~\ref{fig:pptot}, the model
would not be able to accomodate both the beginning of
the rise and the more asymptotic rise at high energy. The origin of
this difficulty needs to be clarified. Here we notice that 
for the 
photon case the behaviour of the 
data beyond the 100 GeV range is not yet clear from the data, given the 
large uncertainties involved and, at present, extrapolation from 
$\gamma \ p$ total cross-section appears not to be too much
off the mark for the present experimental results. Some of the
uncertainties of the models feed into the MonteCarlo
simulation, and it appears that only a dedicated experiment, 
like possibly at the Linear Collider~\cite{LC}, 
can resolve the differences and shed final light on QCD inputs. We show, 
in the two tables,
a compilation of cross-sections at various c.m. $\gamma \gamma$ energies and 
what experimental precision would be required in order to distinguish 
among them. If the difference among the  models indicated, has to be 
more than one standard deviation, then the precision required has to be
the one indicated in the last column in each table.

 In Table ~\ref{tab:table1} we show total $\gamma \gamma$ cross-sections for the three 
 models indicated. The last column shows the 1$\sigma$ level 
 precision needed to discriminate between Aspen~\cite{aspen}
 and T.T. Wu~\cite{ttwu} models. The 
difference between DL~\cite{DL} and either Aspen or T.T. Wu is bigger 
than between 
Aspen and T.T. Wu at each energy value.
\begin{table}
\begin{center}
\caption{\label{tab:table1}Total $\gamma \gamma $ cross-sections and 
required precision for
models based on factorization}
\begin{tabular}{|c||c|c|c|c||}
\hline 
$\sqrt{s_{\gamma \gamma}} (GeV)$ & Aspen &  T.T. Wu & DL & $1 \sigma$ \\ \hline
\hline
 20    & 309 nb & 330 nb & 379 nb &  7\%  \\ \hline
 50    & 330 nb & 368 nb & 430 nb &  11\%   \\ \hline
 100   & 362 nb & 401 nb & 477 nb &  10\%   \\  \hline
 200   & 404 nb & 441 nb & 531 nb &  9\%   \\  \hline
 500   & 474 nb & 515 nb & 612 nb &  8\%   \\  \hline
 700   & 503 nb & 543 nb & 645 nb &  8\%   \\ \hline
\end{tabular}
\end{center}
\end{table}
\begin{table}
\begin{center}
\caption{\label{tab:table2} 
As in Table \protect{~\ref{tab:table1}} for 
Eikonal Minijet
Models}
\begin{tabular}{|c||c|c|c|c||}
\hline 
$\sqrt{s_{\gamma \gamma}} (GeV)$ & BN,GRV & IPT,GRS & IPT,GRV & $1 \sigma$ \\ 
& ($p_{tmin}$=2 GeV)& ($p_{tmin}$=1.5 GeV)  &  ($p_{tmin}$=2 GeV) & \\ \hline
\hline
 19    & 329 nb & 334 nb & 330 nb &  0.3\%  \\ \hline
 54    & 367 nb & 377 nb & 381 nb &  1\%   \\ \hline
 120   & 454 nb & 473 nb & 513 nb &  4\%   \\  \hline
 204   & 547 nb & 590 nb & 683 nb &  8\%   \\  \hline
 452   & 730 nb & 876 nb & 1098 nb &  18\%   \\  \hline
 767   & 873 nb & 1146 nb & 1477 nb &  27\%   \\ \hline
\end{tabular}
\end{center}
\end{table}
In Table~\ref{tab:table2} we show total $\gamma \gamma$ cross-sections for 
the three different predictions from the EMM~\cite{emm} model.
The label BN refers to an extension of the Soft Gluon Summation
model~\cite{bn2} whereas   the label IPT refer to the `intrinsic 
transverse momentum' ansatz  used in the EMM model described in Ref.~[18].
The various acronyms GRS/GRV indicate the parton densities used.
In the last column we show the 1$\sigma$ lebel precision neded at each 
energy to discriminate between the two  closer values for each energy value.
We should also add here that this also gives an estimate of the uncertainties
in our knowledge of the $\gamma \gamma$ induced hadronic backgrounds at the
LC.

\section*{Acknowledgement}
We thank  Martin Block for many clarifying discussions and Albert De Roeck for
useful suggestions.

\end{document}